%% file: article.tex
\documentclass[12pt]{article}
\usepackage{epsfig}

\input econfmacros.tex
\def\Title#1{\begin{center} {\Large {\bf #1} } \end{center}}

\begin{document}

\Title{Recent results of charmed baryon decays at Belle }



\begin{center}
Bilas Pal\footnote{palbs@ucmail.uc.edu}, University of Cincinnati\\
On behalf of the Belle collaboration\\
\end{center}
\begin{abstract}
  \noindent
We report the recent results of charmed baryon decays, based on the data collected by the Belle experiment at the KEKB  collider. This includes the 
observation of the doubly Cabibbo-suppressed  decay $\Lambda_c^+\to\pi^-K^+p$, search for the decay $\Lambda_c^+\to\phi p \pi^0$, and the branching fraction measurement of 
$\Lambda_c^+\to\pi^+K^-p\pi^0$.

\end{abstract}

\section{Introduction}
In this report, we present the recent results of charmed baryon decays  based on the  data, collected by the Belle experiment at the KEKB $e^+e^-$ asymmetric-energy collider~\cite{KEKB}. 
(Throughout this paper charge-conjugate modes are implied.) 
The experiment
took data at center-of-mass energies corresponding
to several $\Upsilon(nS)$ resonances; the total data sample
recorded exceeds $1~{\rm ab}^{-1}$.

The Belle detector is a large-solid-angle magnetic
spectrometer that consists of a silicon vertex detector
(SVD), a 50-layer central drift chamber (CDC),
an array of aerogel threshold Cherenkov counters
(ACC), a barrel-like arrangement of time-of-flight
scintillation counters (TOF), and an electromagnetic
calorimeter comprised of CsI(Tl) crystals
(ECL) located inside a super-conducting solenoid
coil that provides a 1.5 T magnetic field. An iron
flux-return located outside of the coil is instrumented
to detect $K^0_L$
mesons and to identify muons
(KLM). The detector is described in detail elsewhere~\cite{Belle, svd2}.

\section{Observation of the doubly Cabibbo-suppressed \mbox{\boldmath$\Lambda_c^+$} decay }
Several doubly Cabibbo-suppressed (DCS) decays of charmed mesons have been observed~\cite{Agashe:2014kda}. Their measured branching ratios with respect to the corresponding Cabibbo-favored (CF) decays play an important role in
constraining models of the decay of charmed hadrons and in the study of flavor- $SU(3)$ symmetry~\cite{Lipkin:2002za, Gao:2006nb}. On the other hand, because of the smaller production cross-sections for charmed baryons, DCS decays of charmed
baryons have not yet been observed, and only an upper limit, $\frac{\mathcal{B}(\Lambda_c^+\to pK^+\pi^-)}{\mathcal{B}(\Lambda_c^+\to pK^-\pi^+)}<0.46\%$ at 90\% confidence level (CL), has been reported by the FOCUS Collaboration~\cite{Link:2005ym}. Here we present the first observation of the DCS decay $\Lambda_c^+\to pK^+\pi^-$ and the measurement of its branching ratio with respect to the CF decay $\Lambda_c^+\to pK^-\pi^+$, using $980~{\rm fb}^{-1}$ of data~\cite{Yang:2015ytm}.

\begin{figure}[htb!]
\begin{center}
\includegraphics[width=\textwidth, height=6.35cm]{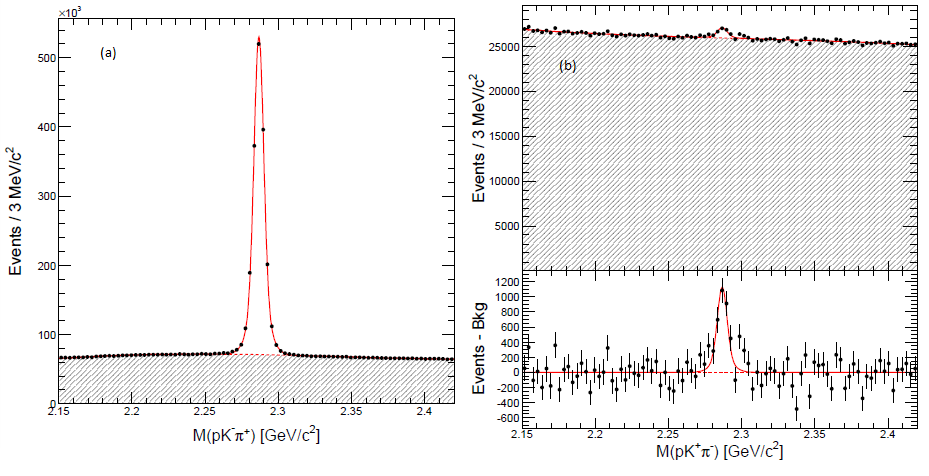}
\vskip -0.3cm
\caption{\small Distributions of (a) $M(pK^-\pi^+)$ and (b) $M(pK^+\pi^-)$ and residuals of data with respect to the fitted combinatorial background. The solid curves indicate the full fit model  and the dashed curves the combinatorial background.}
\label{fig:dcs}
\end{center}
\end{figure}

Figure~\ref{fig:dcs} shows invariant mass distributions of (a) $pK^-\pi^+$ (CF) and (b) $pK^+\pi^-$ (DCS) combinations. DCS decay events are clearly observed in $M(pK^+\pi^-)$. In order to obtain the signal yield, a binned least-$\chi^2$ fit is performed. From the mass fit, we extract $(1.452\pm0.015)\times10^6$ $\Lambda_c^+\to pK^-\pi^+$ events and $3587\pm380$ $\Lambda_c^+\to pK^+\pi^-$ events. The latter has a peaking background from the single Cabibbo-suppressed (SCS) decay $\Lambda_c^+\to\Lambda(\to p\pi^-)K^+$, which has the same final-state topology. After subtracting the SCS contribution, we have $3379\pm380\pm78$ DCS events, where the first uncertainty is statistical and the second is the systematic due to SCS subtraction. The corresponding statistical significance is 9.4 standard deviations. We measure the branching ratio,
\begin{center}
$\frac{\mathcal{B}(\Lambda_c^+\to pK^+\pi^-)}{\mathcal{B}(\Lambda_c^+\to pK^-\pi^+)}=(2.35\pm0.27\pm0.21)\times10^{-3}$, 
\end{center}
where the uncertainties are statistical and systematic, respectively.  This measured branching ratio corresponds to $(0.82\pm0.21)\tan^4\theta_c$, where the uncertainty is the total, which is consistent  within 1.5 standard deviations with the 
na{\"i}ve expectation ($\sim\tan^4\theta_c$~\cite{Link:2005ym}). Multiplying this ratio with the previously measured $\mathcal{B}(\Lambda_c^+\to pK^-\pi^+)=(6.84\pm0.24^{+0.21}_{-0.27})\%$ by the Belle Collaboration~\cite{Zupanc:2013iki}, we obtain the 
 the absolute branching fraction of the DCS decay, 
 \begin{center}
 $\mathcal{B}(\Lambda_c^+\to pK^+\pi^-)=(1.61\pm0.23^{+0.07}_{-0.08})\times10^{-4}$,
 \end{center}
where the first uncertainty is the total uncertainty of the branching ratio and the second is uncertainty of the branching fraction of the CF decay. 
After subtracting the contributions of $\Lambda^*(1520)$ and $\Delta$ isobar intermediates, which contribute only to the CF decay, the revised ratio,  $\frac{\mathcal{B}(\Lambda_c^+\to pK^+\pi^-)}{\mathcal{B}(\Lambda_c^+\to pK^-\pi^+)}=(1.10\pm0.17)\tan^4\theta_c$ is consistent with the na{\"i}ve expectation within 1.0 standard deviation.
\section{Search for \mbox{\boldmath$\Lambda_c^+\to\phi p \pi^0$} and branching fraction measurement of  \mbox{\boldmath$\Lambda_c^+\to K^-\pi^+ p \pi^0$} }

\begin{figure}[htb]
\centering
\includegraphics[width=0.35\textwidth]{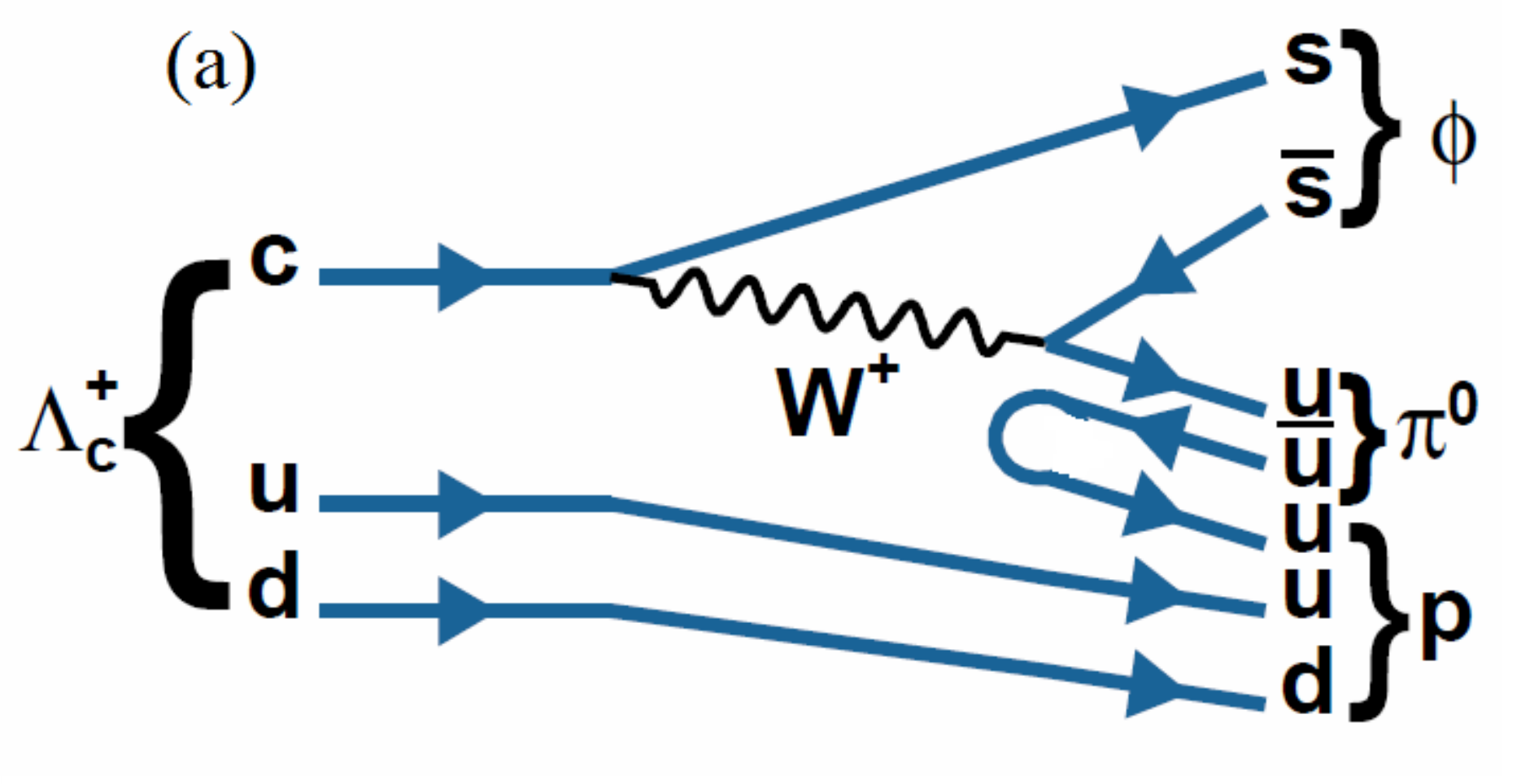}%
\includegraphics[width=0.35\textwidth]{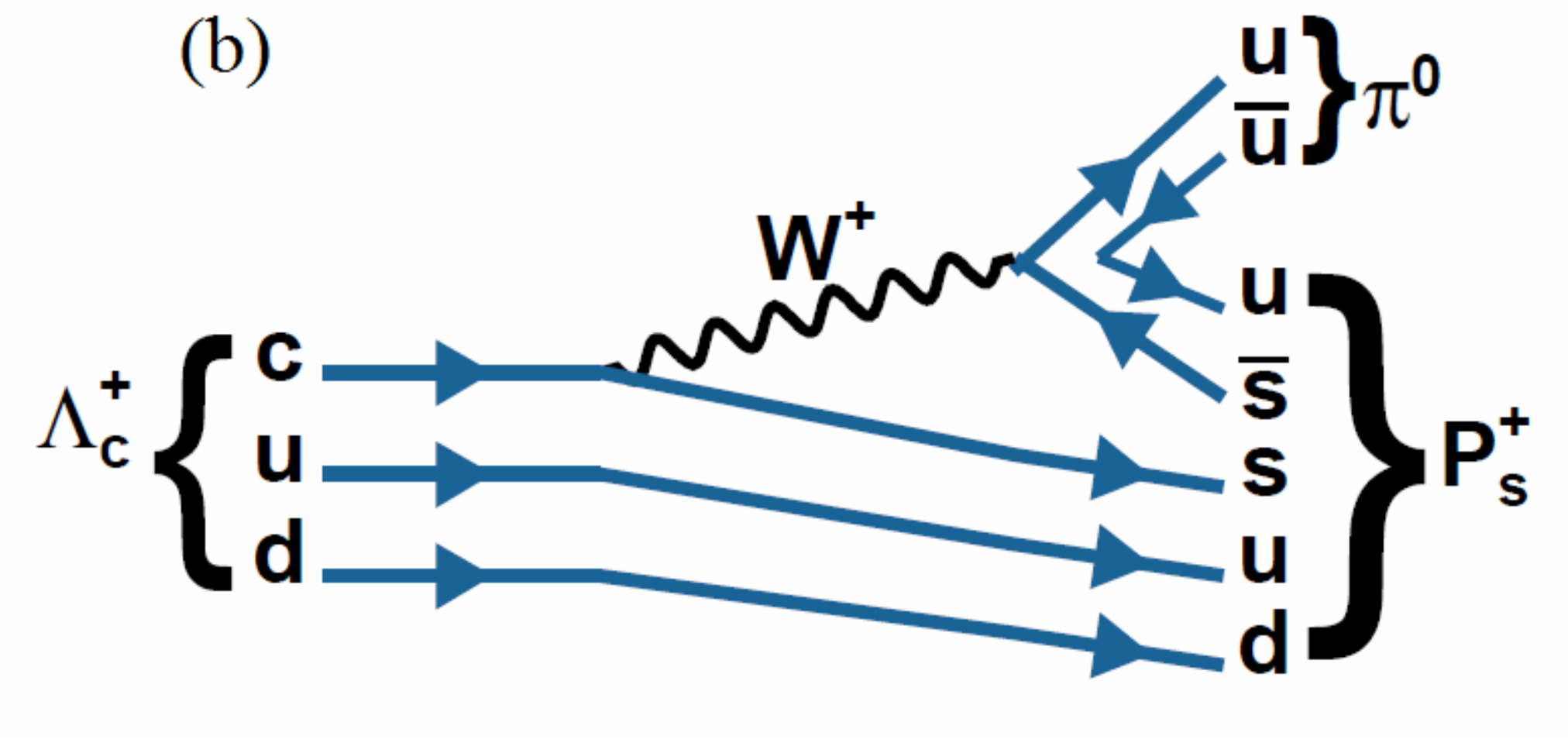}
\vskip -0.3cm
\caption{\small Feynman diagram for the decay (a) $\Lambda_c^+\to\phi p\pi^0$ and (b) $\Lambda_c^+\to P_s^+\pi^0$.}
\label{fig:Feynman}
\end{figure}

The story of exotic hadron  spectroscopy   begins  with the  discovery  of  the
$X(3872)$ by the Belle collaboration in 2003~\cite{Choi:2003ue}. Since then, many exotic $X\!Y\!Z$ states have been reported by Belle and other experiments~\cite{Agashe:2014kda}. Recent observations of two hidden-charm pentaquark  states $P_c^+(4380)$ and $P_c^+(4450)$ by the LHCb collaboration in the $J/\psi p$ invariant mass spectrum of the  $\Lambda_b^0\to J/\psi pK^- $ process~\cite{Aaij:2015tga} raises the question of whether a hidden-strangeness pentaquark  $P_s^+$, where the $c\bar{c}$ pair  in  $P_c^+$  is replaced by  an $s\bar{s}$ pair, exists~\cite{Kopeliovich:2015vqa, Zhu:2015bba, Lebed:2015dca}. The strange-flavor analogue of the $P_c^+$ discovery channel is the decay $\Lambda_c^+\to\phi p\pi^0$~\cite{Kopeliovich:2015vqa, Lebed:2015dca}, shown in Fig.~\ref{fig:Feynman} (a). The detection of a hidden-strangeness pentaquark could be possible through  the $\phi p$ invariant mass spectrum within this channel [see Fig.~\ref{fig:Feynman} (b)]
if the underlying mechanism creating the $P_c^+$ states also holds for $P_s^+$, independent of the flavor~\cite{Lebed:2015dca}, and only if  the mass of $P_s^+$ is less than $M_{\Lambda_c^+}-M_{\pi^0}$.  
In an analogous $s\bar{s}$ process of $\phi$ photoproduction $(\gamma p\to\phi p)$,  a forward-angle  bump structure at $\sqrt{s}\approx2.0$ GeV 
has been observed by  the LEPS~\cite{Mibe:2005er} and CLAS collaborations~\cite{Dey:2014tfa}.
However, this structure appears only at the most forward angles, 
which is  not   expected for  the decay of a resonance~\cite{Lebed:2015fpa}.

Previously, the decay $\Lambda_c^+\to\phi p\pi^0$ has not been studied by any experiment. Here, we report a search for this decay, using 915 $\rm fb^{-1}$ of data~\cite{Pal:2017ypp}. In addition, we search for the nonresonant decay $\Lambda_c^+\to K^+K^-p\pi^0$ and measure the branching fraction of the  Cabibbo-favored decay $\Lambda_c^+\to K^-\pi^+p\pi^0$.

\begin{figure*}[h!tp]
\begin{center}
    \includegraphics[width=0.45\textwidth]{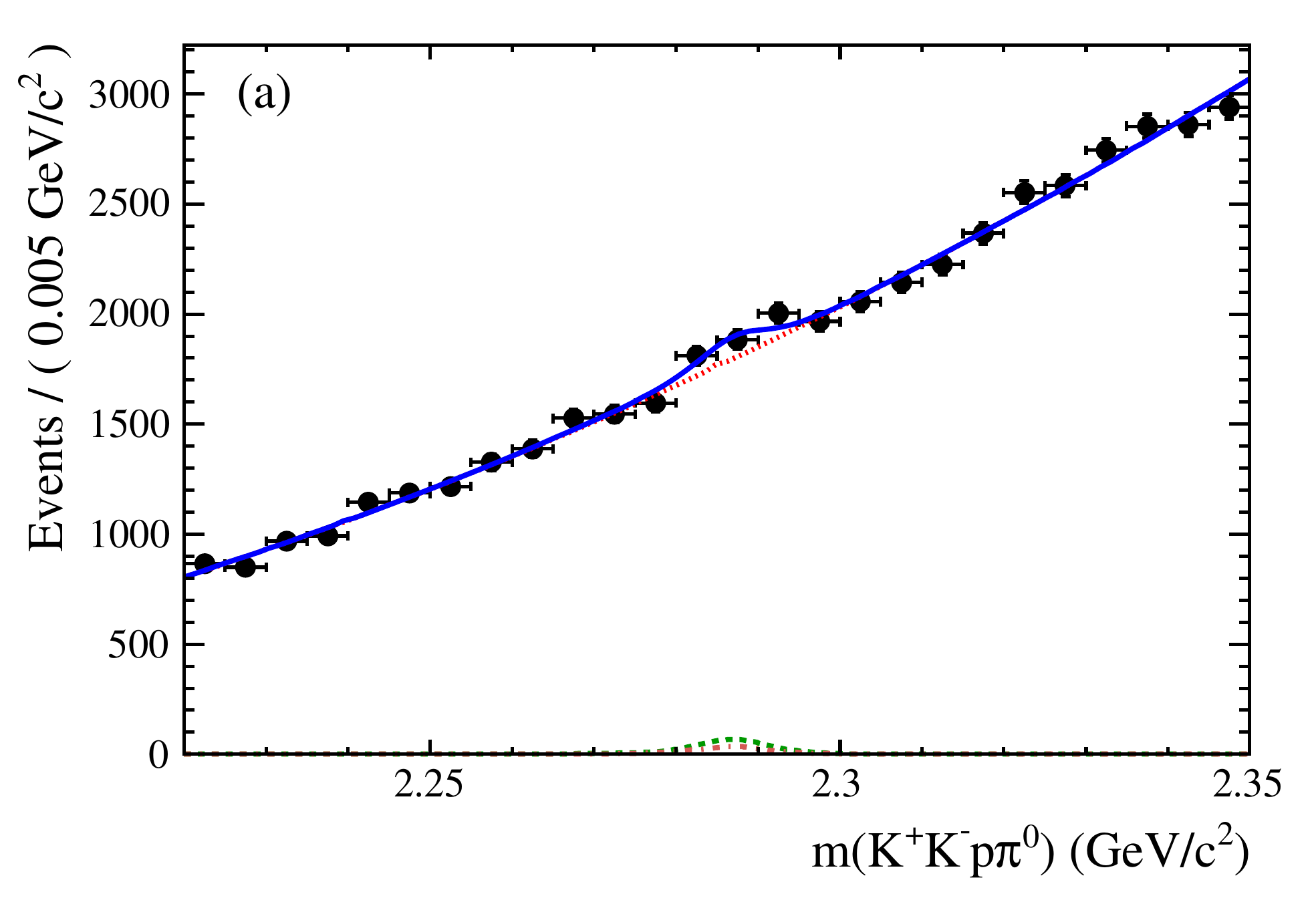}%
       \includegraphics[width=0.45\textwidth]{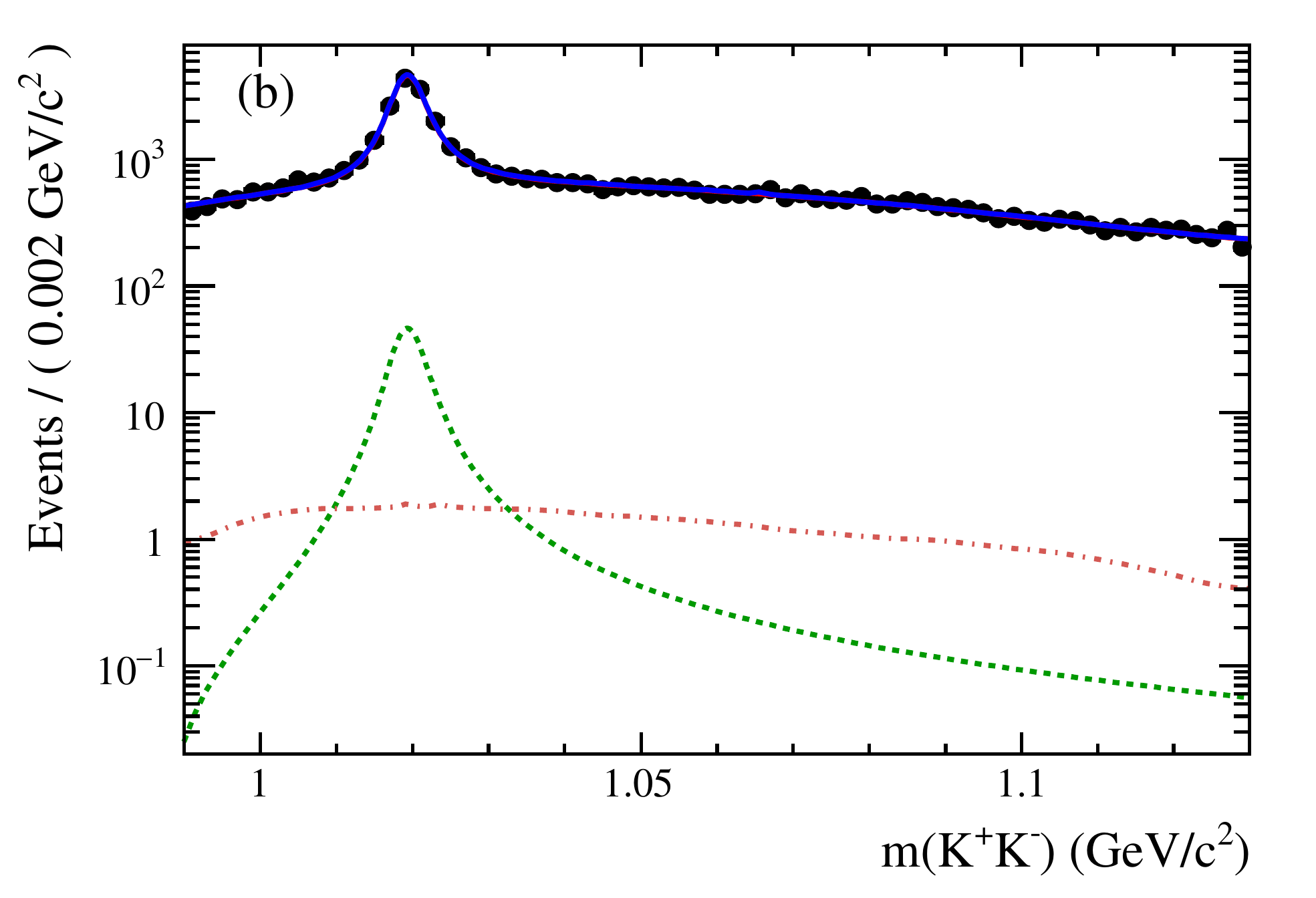}
\end{center}
\vskip -0.75cm
\caption{\small Projections of the 2D fit: (a) $m(K^+K^- p\pi^0)$ and (b) $m(K^+K^-)$. The points with the error bars are the  data, and the (red) dotted, (green) dashed and (brown) dot-dashed curves represent the combinatorial,  signal and nonresonant candidates, respectively, and (blue) solid curves represent the total PDF. The solid curve in (b) completely overlaps the curve for the combinatorial background.}
\label{fig:2dfit}
\end{figure*}

In order to extract the signal yield, we perform a two-dimensional (2D) unbinned extended maximum likelihood fit to the variables $m (K^+K^-p\pi^0)$ and  $m(K^+K^-)$.  Projections of the fit result are shown in Fig.~\ref{fig:2dfit}. 
From the fit, we extract $148.4\pm61.8$ signal events, $75.9\pm84.8$ nonresonant events, 
and $7158.4\pm36.4$ combinatorial background events. 
The statistical significances are found to be 2.4 and 1.0 standard deviations for $\Lambda_c^+\to\phi p \pi^0$ and nonresonant $\Lambda_c^+\to K^+K^- p \pi^0$ decays, respectively. We use the well-established decay $\Lambda_c^+\to p K^-\pi^+$~\cite{Agashe:2014kda} as the normalization channel for the branching fraction measurements. The branching fraction is calculated as
\begin{eqnarray}
\mathcal{B}(\Lambda_c^+\to {\rm final~state}) &=& \frac{Y_{\rm Sig}/\varepsilon_{\rm Sig}}{Y_{\rm Norm}/\varepsilon_{\rm Norm}}\nonumber \\ 
&\times& \mathcal{B}(\Lambda_c^+\to pK^-\pi^+) ,\label{eq:br}
\end{eqnarray}
where $Y$ represents the observed yield in the signal region of the decay of interest and $\varepsilon$ corresponds to the reconstruction efficiency as obtained from the MC simulation,  and $\mathcal{B}(\Lambda_c^+\to pK^-\pi^+) =(6.46\,\pm0.24)\%$~\cite{Amhis:2016xyh}. For the $\phi p\pi^0$ final state, we include  $\mathcal{B}(\phi\to K^+K^-)=(48.9\,\pm0.5)\%$~\cite{Agashe:2014kda} in the denominator of Eq.~(\ref{eq:br}).

Since the significances are below 3.0 standard deviations both for $\phi p\pi^0$ signal and $K^+K^-p\pi^0$ nonresonant decays, we set upper limits on their branching fractions  at 90\% CL using a Bayesian approach. The limit is obtained by integrating the
likelihood function from zero to infinity; the value that
corresponds to 90\% of this total area is taken as the
90\% CL upper limit.  We include the systematic uncertainty in the calculation by
convolving the likelihood distribution with a Gaussian function
whose width is set equal to the total systematic uncertainty. The results are
\begin{eqnarray*}
\mathcal{B}(\Lambda_c^+\to \phi p\pi^0) &<& 15.3\times10^{-5} ,\\
\mathcal{B}(\Lambda_c^+\to K^+K^-p\pi^0)_{\rm NR} &<&6.3\times10^{-5} ,
\end{eqnarray*}
which are the first limits on these branching fractions. 

\begin{figure}[htb]
\centering
\includegraphics[width=0.5\textwidth]{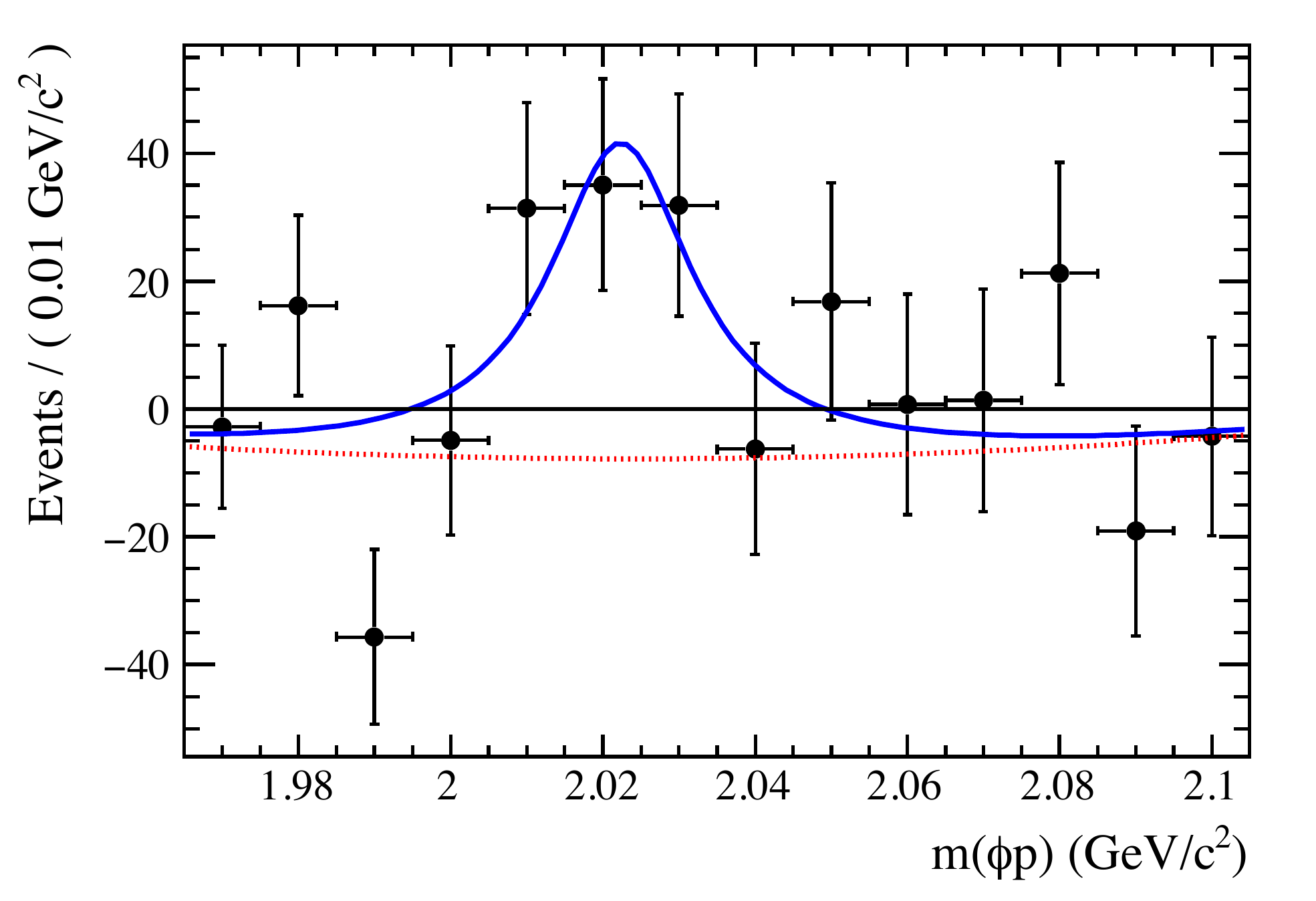}
\vskip -0.3cm
\caption{\small   The background-subtracted distribution of $m(\phi p)$ in the $\phi p\pi^0$ final state. The points with error bars are data, and   the (blue) solid line shows the total PDF. The (red) dotted curve shows the fitted phase space component (which has fluctuated negative).}
\label{fig:bkg-sub_dis}
\end{figure}

To search for a putative $P_s^+\to\phi p$ decay, we select $\Lambda_c^+\to K^+K^-p\pi^0$ candidates in which $m(K^+K^-)$ is within 0.020~GeV/$c^2$ of the   $\phi$ meson mass~\cite{Agashe:2014kda}
and plot the  background-subtracted $m(\phi p)$ distribution (Fig.~\ref{fig:bkg-sub_dis}). This distribution is obtained by performing 2D fits as discussed above in bins of $m(\phi p)$. 
The data shows no clear evidence for a $P_s^+$ state. 
We set an upper limit on the product branching fraction 
$\mathcal{B}(\Lambda_c^+\to P_s^+\pi^0) \times \mathcal{B}(P_s^+\to \phi p)$ by fitting the  distribution of Fig.~\ref{fig:bkg-sub_dis}
to the sum of a RBW function and a phase space 
distribution determined from a sample of simulated $\Lambda^+_c\to\phi p\pi^0$ 
decays. We obtain $77.6\pm28.1$ $P_s^+$ events from the fit, which gives an upper limit of 
\begin{center}
$\mathcal{B}(\Lambda_c^+\to P_s^+\pi^0) \times 
\mathcal{B}(P_s^+\to \phi p)  <  8.3\times 10^{-5}$ 
\end{center}
at 90\% CL. This limit is calculated using the same procedure as that used for our limit on ${\cal B}(\Lambda_c^+\rightarrow \phi p \pi^0)$. From the fit, we also obtain, 
\begin{center}
$M_{P_s^+}=(2.025\pm 0.005)$~GeV/$c^2$ and 
$\Gamma_{P_s^+}=(0.022\pm 0.012)$~GeV,
\end{center}
where the uncertainties are statistical only.

\begin{figure}[h!tb]
\centering
\includegraphics[width=0.5\textwidth]{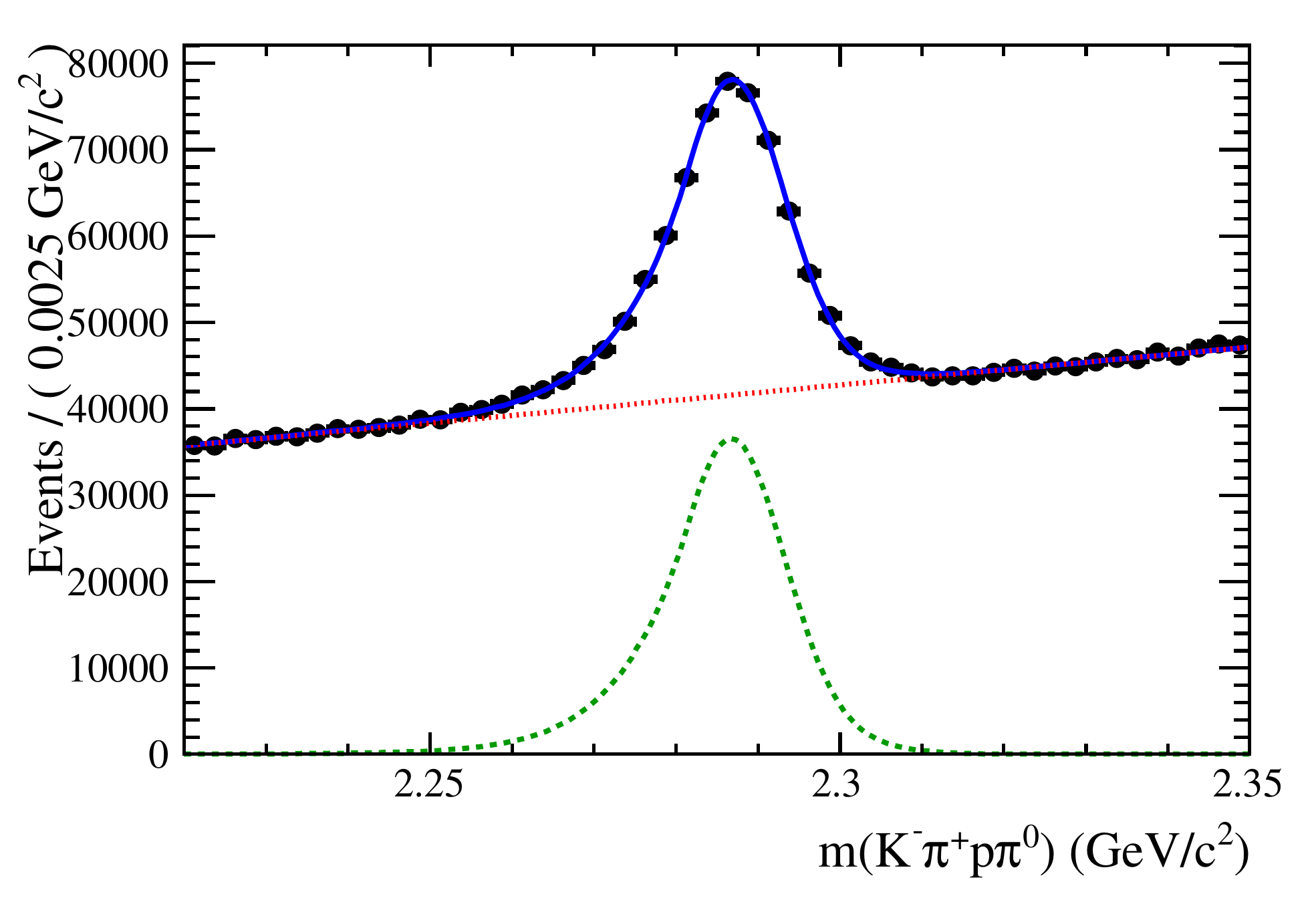}
\vskip -0.3cm
\caption{\small  Fit to the invariant mass distribution of $m(K^-\pi^+p\pi^0)$. The points with the error bars are the  data,  the (red) dotted and (green) dashed curves  represent the combinatorial and  signal  candidates, respectively, and (blue) curve represents the total PDF.  The $\chi^2/$ (number of bins) of the fit is 1.43, which indicate that the fit gives a good description of the data.}
\label{fig:invmass_control2_data}
\end{figure}

The high statistics decay  $\Lambda_c^+\to K^-\pi^+p\pi^0$ is used to adjust the data-MC differences in the $\phi p\pi^0$ signal and $K^+K^-p\pi^0$ nonresonant decays. For the $\Lambda_c^+\to K^-\pi^+p\pi^0$ sample, 
the mass distribution is plotted in Fig.~\ref{fig:invmass_control2_data}. We fit this distribution to obtain the signal yield. We find $242\,039\pm \,2342$ signal candidates and $472\,729\pm\,467$ background candidates. We measure the ratio of branching fractions,
\begin{center}
$\frac{\mathcal{B}(\Lambda_c^+\to K^-\pi^+p\pi^0)}{\mathcal{B}(\Lambda_c^+\to K^-\pi^+p)}=(0.685\pm0.007\pm 0.018),$
\end{center}
where the first uncertainty is statistical and the second is systematic. Multiplying this ratio by the world average value of $\mathcal{B}(\Lambda_c^+\to K^-\pi^+p)=(6.46\pm0.24)\%$~\cite{Amhis:2016xyh}, we obtain
\begin{center}
$\mathcal{B}(\Lambda_c^+\to K^-\pi^+p\pi^0)=(4.42\pm0.05\pm 0.12\pm0.16)\%,$
\end{center}
where the first uncertainty is statistical, the second is systematic, and the third reflects the uncertainty due to the branching fraction of the normalization decay mode ($\mathcal{B}_{\rm Norm}$). This is the most precise measurement of $\mathcal{B}(\Lambda_c^+\to K^-\pi^+p\pi^0)$ to date and is consistent with the recently measured value $\mathcal{B}(\Lambda_c^+\to K^-\pi^+p\pi^0)=(4.53\pm0.23\pm0.30)\%$ by the BESIII collaboration~\cite{Ablikim:2015flg}.

\section*{Acknowledgements}
\noindent
The author thanks the organizers of DPF 2017 for  excellent hospitality and for assembling a nice scientific program. This work is supported by the U.S. Department of Energy.

\end{document}

%% file: econfmacros.tex
\def\beq{\begin{equation}}
\def\eeq#1{\label{#1}\end{equation}}
\def\eeqn{\end{equation}}

\def\beqa{\begin{eqnarray}}
\def\eeqa#1{\label{#1}\end{eqnarray}}
\def\eeqan{\end{eqnarray}}

\let\bar=\overbar

\def\Dslash{\not{\hbox{\kern-4pt $D$}}}
\def\dslash{\not{\hbox{\kern-2pt $\del$}}}

\def\msb{{\bar{\ssstyle M \kern -1pt S}}}

